\documentclass[letterpaper, 10 pt, conference]{IEEEtran}  % Comment this line out if you need a4paper

\usepackage{graphics} % for pdf, bitmapped graphics files
\usepackage{graphicx}

\usepackage{url}

\usepackage{subcaption}
\usepackage{comment} % enables the use of multi-line comments (\ifx \fi) 
\usepackage{color}
\usepackage[dvipsnames,svgnames,x11names]{xcolor}
\usepackage[markup=underlined]{changes}

\definechangesauthor[color=RedViolet]{N}
\definechangesauthor[color=OliveGreen]{A}
\definechangesauthor[color=NavyBlue]{V}

\usepackage{amsmath} % assumes amsmath package installed
\usepackage{amssymb}  % assumes amsmath package installed

\usepackage{cite}
\usepackage[dvipsnames]{xcolor}

\usepackage{comment}

\newif\ifshow
\showtrue
\ifshow

\else
\excludecomment{scratch}
\fi

\title{\LARGE \bf Deep Predictive Coding Neural Network for RF Anomaly Detection in Wireless Networks}

\author{Nistha Tandiya, Ahmad Jauhar, Vuk Marojevic, Jeffrey H. Reed \\
Wireless@Virginia Tech, Bradley Department of Electrical and Computer Engineering,\\
Virginia Tech, Blacksburg, VA, USA. \\
Email: \{nistha, ahmadsj, maroje, reedjh\}@vt.edu% <-this % stops a space
\thanks{
This is the author’s version of the work. For citation purposes, the definitive Version of Record of this work is: N. Tandiya, A. Jauhar, V. Marojevic, J.H. Reed, ``Deep Predictive Coding Neural Network for RF Anomaly Detection in Wireless Networks,'' in \textit{Communications Workshops (ICC Workshops), 2018 IEEE International Conference}, 20-24 May 2018, to appear on
IEEE Xplore digital library

Personal use of this material is permitted. Permission from IEEE must be obtained for all other uses, in any current or future media, including 
reprinting/republishing this material for advertising or promotional purposes, creating new collective works, for resale or redistribution to servers or lists, or reuse of any copyrighted component of this work in other works.
}
}

\begin{document}
% Copyright information for Author's online redistribution
\IEEEoverridecommandlockouts
\IEEEpubid{\copyright{} Copyright 2018 IEEE}
%\pagenumbering{gobble}
\maketitle
%\thispagestyle{plain}
%\pagestyle{plain}

%%%%%%%%%%%%%%%%%%%%%%%%%%%%%%%%%%%%%%%%%%%%%%%%%%%%%%%%%%%%%%%%%%%%%%%%%%%%%%%%
\begin{abstract}
%Security is one of the biggest hurdles for next-generation wireless networks to be able to reliably support mission-critical as well as utility services and satisfy people's daily needs. %systems as the current wireless protocols are unable to handle its vulnerabilities. 
Intrusion detection has become one of the most critical tasks in a wireless network to prevent service outages that can take long to fix. The sheer variety of anomalous events necessitates adopting cognitive anomaly detection methods instead of the traditional signature-based detection techniques. This paper proposes an anomaly detection methodology for wireless systems that is based on monitoring and analyzing radio frequency (RF) spectrum activities. Our detection technique leverages an existing solution for the video prediction problem, and uses it on image sequences generated from monitoring the wireless spectrum. The deep predictive coding network is trained with images corresponding to the normal behavior of the system, and whenever there is an anomaly, its detection is triggered by the deviation between the actual and predicted behavior. For our analysis, we use the images generated from the time-frequency spectrograms and spectral correlation functions of the received RF signal. We test our technique on a dataset which contains anomalies such as jamming, chirping of transmitters, spectrum hijacking, and node failure, and evaluate its performance using standard classifier metrics: detection ratio, and false alarm rate. Simulation results demonstrate that the proposed methodology effectively detects many unforeseen anomalous events in real time. We discuss the applications, which encompass industrial IoT, autonomous vehicle control and mission-critical communications services. \\  
%Specifically, we convert the raw wireless data into time-frequency spectrograms and spectral correlation functions (SCF), and record these as sequential image data. 
%Our approach is based on using a deep neural network video predictor on sequential images generated from the wireless spectrum. Specifically, we convert the raw wireless data into time-frequency spectrograms and spectral correlation functions (SCF), and record these as sequential image data. These image sequences are inputs for the predictor module. 
%The predictor is trained with images corresponding to the normal behavior of the system. Consequently, during operation, it assumes normal functioning of the system to predict future behavior. When an anomaly occurs in the system, there is a deviation between the actual and predicted behavior of the system, and this triggers its detection. 
%We evaluate our approach by testing it with anomalies such as jamming, chirping of transmitters, spectrum hijacking, and node failure. The experimental results illustrate that our method can efficiently detect many unforeseen anomalous events, and can therefore enable real-time automated monitoring of the spectrum.  \\ \\

\end{abstract}
%%%%%%%%%%%%%%%%%%%%%%%%%%%%%%%%%%%%%%%%%%%%%%%%%%%%%%%%%%%%%%%%%%%%%%%%%%%%%%%%
\section{Introduction}\label{sec:intro}
Wireless technology enables many services and applications and will play a key role for enabling the smart and autonomous systems of the future. Wireless networks also provide mission-critical infrastructure for public safety, national security, and military communications. This popularity of wireless technology can be attributed to its ease of access and high availability; however, these very same features contribute to many of its vulnerabilities. Attacks such as jamming, spoofing, and eavesdropping have existed since the early days of analog systems, and despite important advances enabled by digital technology, they continue to exist in today's and emerging wireless networks \cite{marojevic2017performance,lichtman2016lte,mina_icnc,rupprecht2017security,marc2018icc}. The openness of wireless standards and the inherently open wireless channel can be exploited by adversaries in the \textit{reconnaissance} and \textit{exploitation} phases of the cyber kill chain \cite{martin2014cyber} to launch cyber attacks.

Despite continuous progress in managing cyber risks, the development of cyber security measures cannot keep up with the growth rate and diversity of cyber attacks \cite{cornish2010cyber}. Preventive security measures are  ineffective against unforeseen or zero-day attacks. Furthermore, adversaries can surpass preventive measures by finding vulnerabilities associated with design imperfections, implementation errors, or incorrect configurations \cite{yu2011intrusion}. Thus, it is a well accepted fact that the anticipation and prevention of all possible attacks and malfunctions are not feasible for current or future cyber-physical systems \cite{linkov2013resilience}.

% Properly align column margins to display IEEE copyright notice
\IEEEpubidadjcol

Once a system is attacked by an adversary, the first step in its recovery process is to determine the existence of a system anomaly \cite{colbert2016intrusion}. Intrusion detection systems (IDS) are critical components of the security infrastructure and are specifically designed for this purpose. Their primary goal is to collect and analyze system information to detect the presence of malicious activity. This detection is essential because it is the basis for triggering subsequent remediation steps that localize, isolate, and mitigate the threat. However, this initial step is by itself challenging.

The constantly evolving threats and rapid development of new attack patterns render the commercially popular signature-based IDS (e.g. \textit{OpenWIPS-ng}, \textit{Suricata}) ineffective as their detection scheme is only as good as their database of stored signatures. The other alternative---anomaly detection based IDS---are better suited for handling unknown attacks. In order to identify anomalies, they just need to know the “normal” behavior of the system, and whenever a significant deviation from that reference is observed, an intrusion is reported. For wireless networks, the sheer variety of possible anomalous events necessitates adoption of anomaly detection based IDS.

Anomaly detection techniques can be classified based on the layer of the protocol stack from which they derive the data. %; this governs the types of attacks/anomalies the system can handle. % Vuk: don't understand 'governs' and what the message of this subclause
In order to detect attacks such as PHY layer jamming, spectrum hijacking, or MAC address spoofing, an analysis of lower layer attributes---physical (PHY) and data link layer attributes---is required. We are motivated by this requirement and propose a machine learning (ML) based anomaly detection technique which analyzes RF spectral data of a wireless network.

This paper introduces:
\begin{enumerate}
\item An anomaly detection scheme which enables automated and non-intrusive real-time monitoring of RF signals, and
\item A methodology that leverages powerful video prediction tools for anomaly detection in wireless networks.
\end{enumerate}

%Currently, most of the intrusion detection techniques such as \textit{Snort} and \textit{Bro} work at the higher layers of the protocol stack, and are therefore unable to detect attacks such as MAC address spoofing and PHY layer jamming. Detection of these attacks requires analysis of the lower layer attributes---such as PHY and Data Link layer. We are motivated by this requirement, and in this paper, we propose a machine learning (ML) based anomaly detector which analyzes spectral data of a wireless network. Our ML module is an analogue of video predictor PredNet \cite{lotter2016deep}, which predicts spectral behavior of a system. In order to identify anomalous events, our proposed solution uses this predictor's output and compares it to the actual behavior of the network. Whenever the prediction error surpasses a threshold, an anomaly is reported. 

The remainder of this paper is organized as follows: Section~\ref{sec:litsurvey} formulates the problem and summarizes related work. % \added[id=N]{using ML techniques}. 
Section~\ref{sec:method} introduces our methodology. Implementation details and performance evaluation are provided in Section~\ref{sec:results}. We conclude the paper in Section~\ref{sec:conc} with a discussion on applications and an outline for future research.

\section{Problem Formulation and Related work}\label{sec:litsurvey}
Many emerging applications such as smart cities and the smart grid require heavy deployment of wireless devices for their realization. %, making the intrusion detection problem even more challenging. 
Attacks or anomalous behavior in these systems can have severe consequences and their detection requires automated monitoring of huge amounts of data, which can be handled only by ML techniques. The diversity of anomalies in wireless system makes it impossible to obtain their signatures \cite{feng2017anomaly}. Furthermore, IDS for wireless networks have information available only from the physical and the data link layers \cite{alipour2015wireless} as the information present in higher layers are often encrypted and therefore not easily accessible. The problem then consists of developing a methodology for effective signal monitoring and processing, while being adaptable enough to accommodate unknown threats in the future.

Reference \cite{yin2009temporal} proposes a data-mining based approach to identify anomalies in temporal-spectral data. The method creates historical models from the previously recorded data and compares real-time measurements with these models. Our method follows a similar approach, but instead of saving the entire past data, we just train the ML module with a normal-behavior dataset. Thus, our method can be much faster as compared to \cite{yin2009temporal}. The authors of \cite{afgani2010information} propose two anomaly detection methods using information theoretic measures: Kullback-Leibler divergence (KLD) and information content. They statistically analyze the signal envelope to create empirical event probabilities and use it with the two measure functions to identify interference instances. Reference \cite{alipour2015wireless} proposes a detection scheme to identify anomalous behavior in IEEE 802.11 Wireless LAN by analyzing MAC layer frames. The authors create a normal model using $n$ consecutive state-machine transitions during the normal functioning of the network. At runtime, their IDS identifies anomalous sessions based on the similarity between the real-time traffic and the pre-trained normal model. A discrete wavelet transform (DWT) based method of anomaly detection is proposed in \cite{saganowski2016dwt} to identify anomalies in wireless sensor network (WSN) traffic. The network traffic is there preprocessed with \textit{Snort} and the resultant features undergo DWT to decompose the energy into different sub-bands. These energy coefficients of the normal traffic behavior are used as a reference to compare against real-time traffic features in order to identify anomalies. 

The authors of \cite{feng2017anomaly} propose an unsupervised learning approach for RF spectrum-based anomaly detection in wireless communications based on a two-layered autoencoder. They consider a one-class classification problem to recognize a signal with high signal-to-noise ration (SNR). This approach is limited to a specific kind of anomaly and does not address wider classes of anomalies. Similar to our approach, \cite{o2016recurrent} uses 2D image representations of the wide-band time-frequency spectrograms for detection, localization and identification of radio transmissions using convolutional neural networks (CNN). However, their solution does not address the anomaly detection problem.

Reference \cite{o2017spectral} demonstrates the efficacy of recurrent neural network architectures for RF anomaly detection. It analyzes raw spectral data using CNN based Long Short Term Memory (LSTM) networks and characterizes the prediction error as a Gaussian distribution. The approach performs well in structured radio environments by detecting deviations from the structured behavior. We approach the RF anomaly detection problem using a deep predictive coding network, which has shown better results than traditional solutions \cite{lotter2016deep}.

\begin{figure}[ht]
\centering
\includegraphics[trim= 0.0cm 0.0cm 0.4cm 0.0cm, clip=true, width=.5\textwidth]{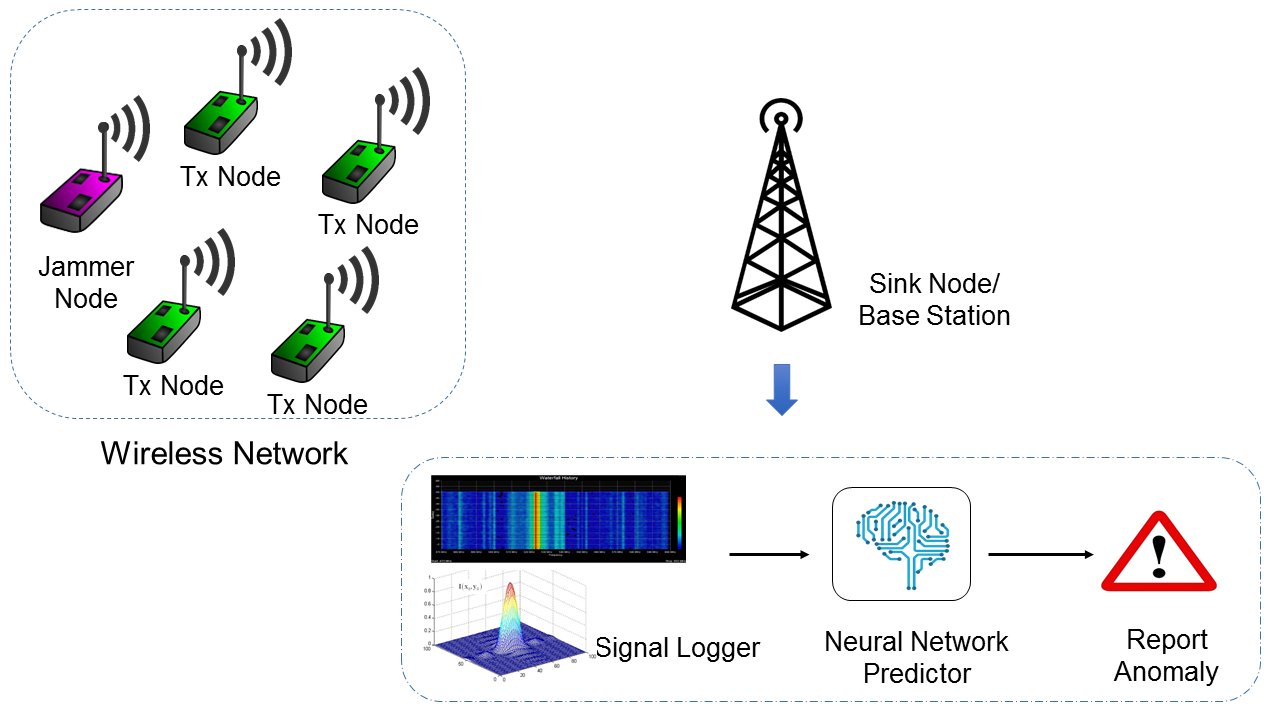}
\caption{\small Block diagram of our system. The proposed anomaly detector sits at the sink node or base station where RF spectrum monitoring is carried out.}
\label{fig:sysModel}
\end{figure}
\begin{figure*}[t!]
\centering
\centering
    \begin{subfigure}[t]{.47\textwidth}
        \centering
       \includegraphics[trim= 0.0cm 9.0cm 0.0cm 0.0cm, clip=true, width=\textwidth]{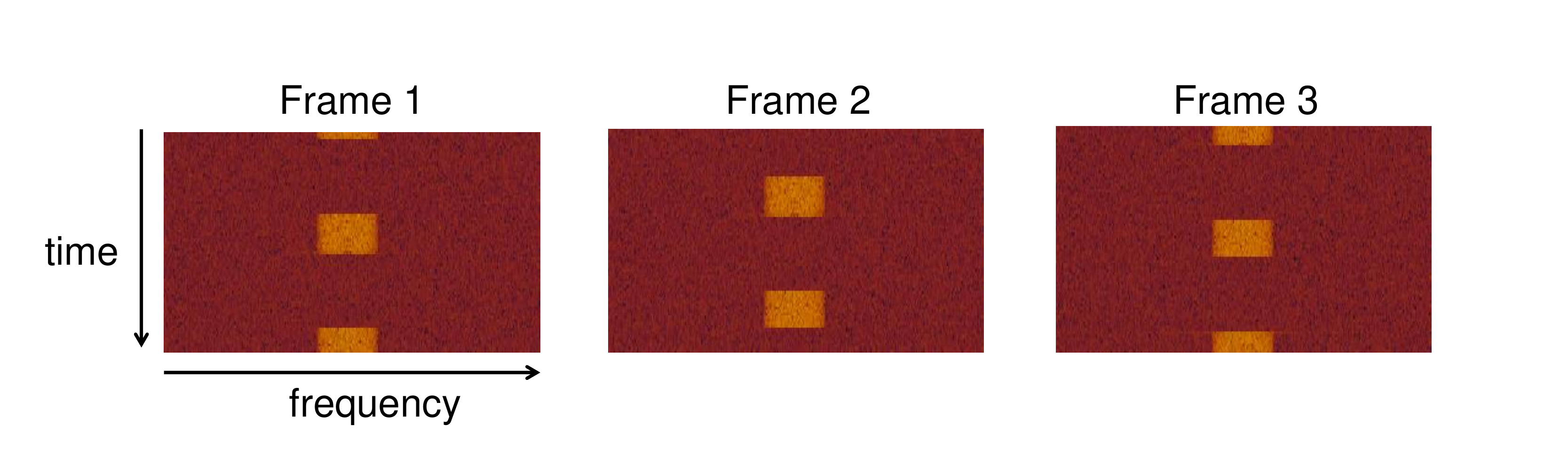}    
        \caption{\small Sequential image frames for the time-frequency spectrogram of a two-node wireless network. Each image captures the spectral behavior over a finite time duration.}
        \label{fig:psd_images}
    \end{subfigure}~     
    \begin{subfigure}[t]{.47\textwidth}
        \centering    
       \includegraphics[trim= 0.0cm 0.0cm 0.0cm 8.0cm, clip=true, width=\textwidth]{./Figures/imageSeq.pdf}    
        \caption{\small Sequential image frames representing the spectral correlation function of the signal of a wireless network. Each frame captures the spectral behavior for a particular time instant.}
        \label{fig:cs_images}
    \end{subfigure}
    \caption{\small Sample input sequence for the video predictor module.}
t    \label{fig:SampleImages}
    \end{figure*} 
    
\section{Methodology}\label{sec:method}
We consider the spectrum monitoring application for the uplink channel of a wireless network as shown in Fig.~\ref{fig:sysModel}. The anomaly detector module is located at a sink node or the base station which monitors the spectrum. %used by the nodes in the network.
In addition to the legitimate nodes, malicious nodes may also be present. At the sink node, the time-domain data is processed in real time and converted to frequency-domain data using the Fast Fourier Transform (FFT) and the Spectral Correlation Function (SCF) before being stored in the form of sequential 2D image representations as shown in Fig.~\ref{fig:SampleImages}.

\begin{figure}
\centering
\includegraphics[trim= 0.0cm 0.0cm 0.0cm 0.0cm, clip=true, width=.3\textwidth]{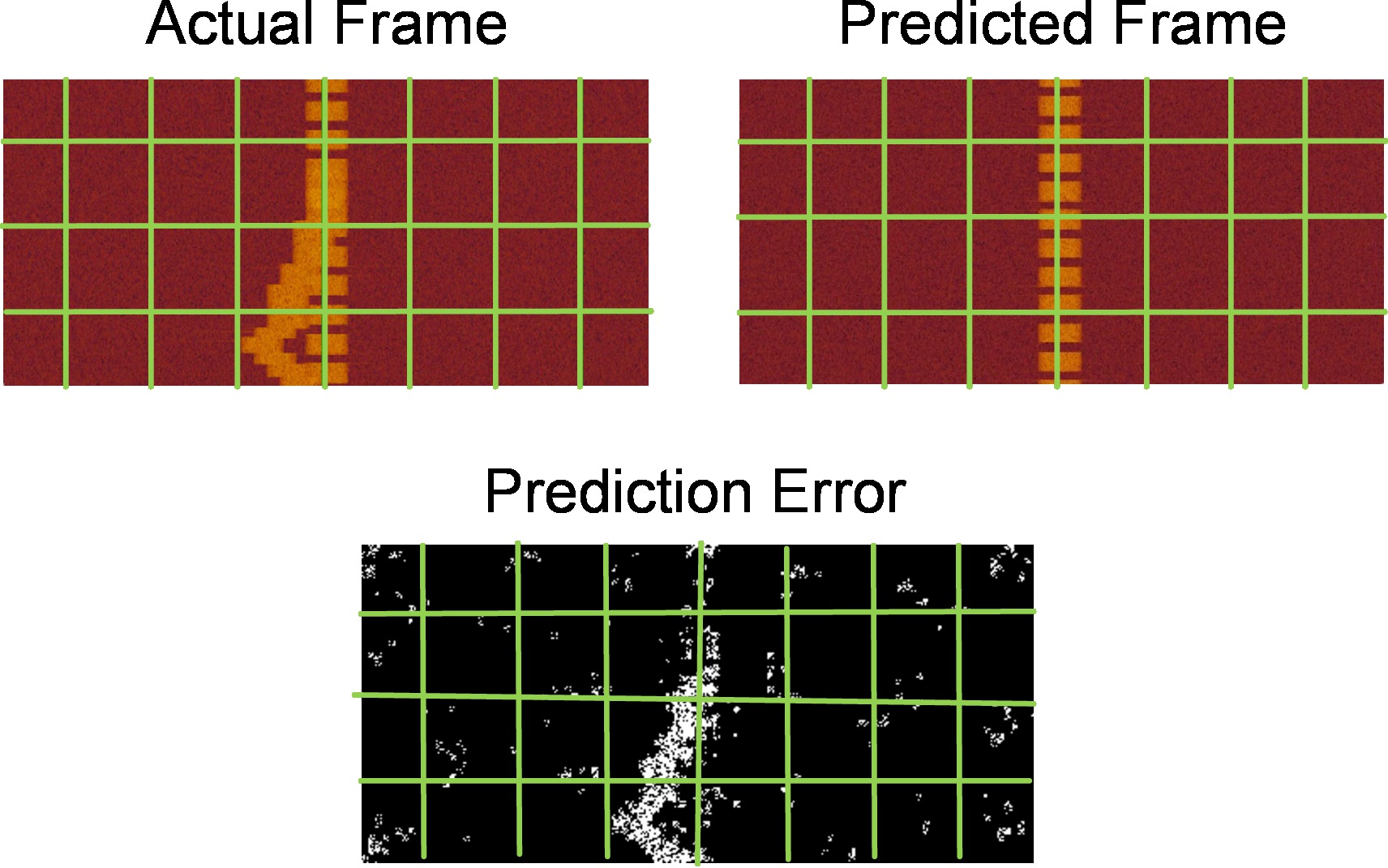}
\caption{\small Division of the spectrogram image into an $m \times n$ grid. Statistics of the absolute difference in the pixel values of the predicted and actual frame are compared for each block.%  with the statistics of the corresponding block of the normal behavior.
}
\label{fig:Grid}
\end{figure}
Our anomaly detector module takes these image sequences as the input and feeds them to a deep learning video predictor, such as \textit{Prednet} \cite{lotter2016deep}, which tries to predict the next frame of a video based on the past frames. We pre-train the weights of the deep predictive coding network with image sequences corresponding to the network's normal operation. This enables the neural network to make predictions and detect anomalies that deviate from normal network conditions. More precisely, the predictor compares the output predictions of each image frame with the image of the actual spectrum behavior using an approach similar to \cite{o2016recurrent}. The comparison is done by segmenting each of the frames into an $m\times n$ grid as shown in Fig.~\ref{fig:Grid} and calculating the mean absolute error for each block of the grid. These error values can be modeled as a stochastic distribution and we use a Gaussian distribution in this paper. During the training process, we correspondingly save the expected error statistics for the normal network behavior to evaluate the likelihood of the mean prediction error for each grid segment. The cumulative likelihood for each of the frames is then used to identify anomalies.

\begin{figure*}[ht]
\centering
	\begin{subfigure}[t]{.2\textwidth}      
    \centering
       \includegraphics[trim= 0.0cm 0.0cm 0.0cm 0.0cm, clip=true, width=\textwidth, height=2.4cm]{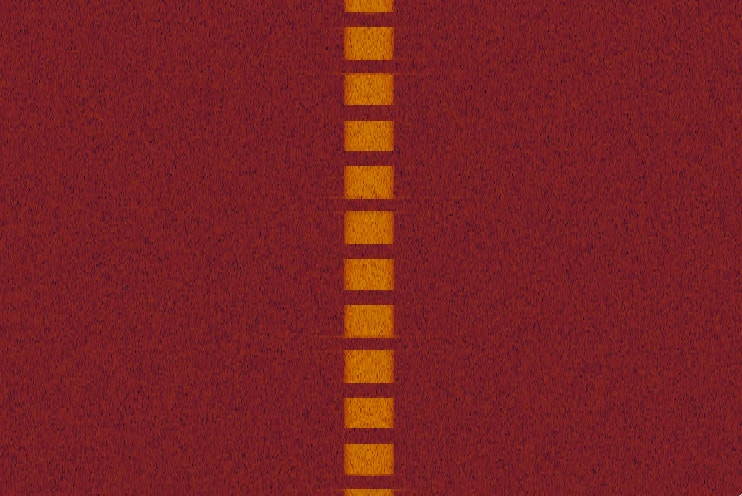}    
       \caption{\small Normal behavior}
    \end{subfigure}~
        \begin{subfigure}[t]{.2\textwidth}      
        \centering
       \includegraphics[trim= 0.0cm 0.0cm 0.0cm 0.0cm, clip=true, width=\textwidth, height=2.4cm]{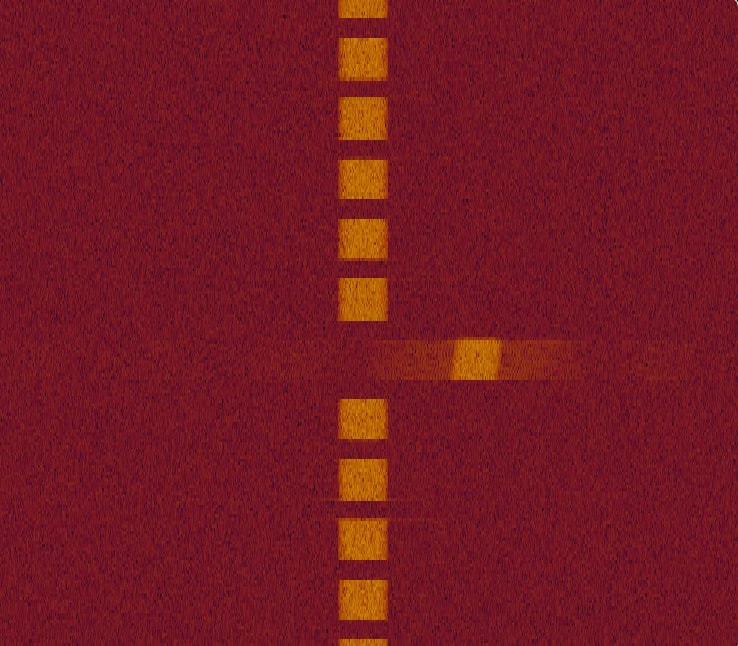}    
       \caption{\small Single chirping node}
        \end{subfigure}~
        \begin{subfigure}[t]{.2\textwidth}      
        \centering
       \includegraphics[trim= 0.0cm 0.0cm 0.0cm 0.0cm, clip=true, width=\textwidth, height=2.4cm]{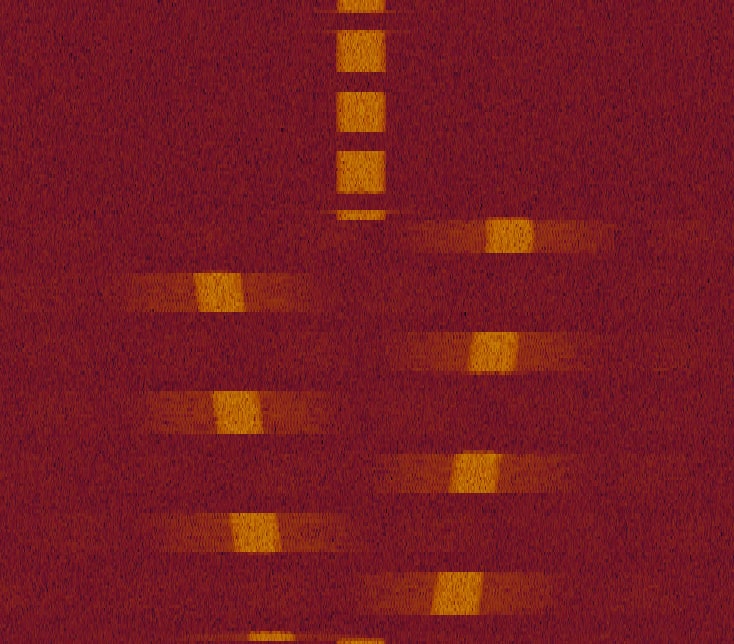}    
       \caption{\small Two nodes chirping}
        \end{subfigure}~
        \begin{subfigure}[t]{.2\textwidth}      
        \centering
       \includegraphics[trim= 0.0cm 0.0cm 0.0cm 0.0cm, clip=true, width=\textwidth, height=2.4cm]{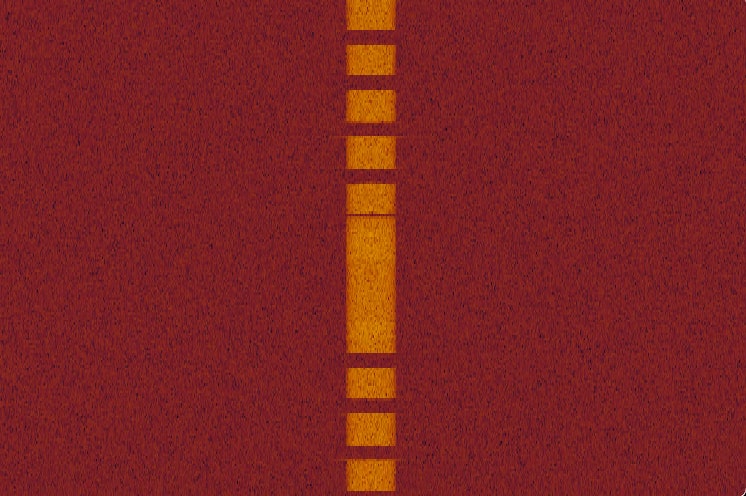}    
       \caption{\small Spectrum hijacking}
        \end{subfigure}
                        
               \begin{subfigure}[t]{.2\textwidth}     
        \centering
       \includegraphics[trim= 0.0cm 0.0cm 0.0cm 0.0cm, clip=true,  height=2.4cm]{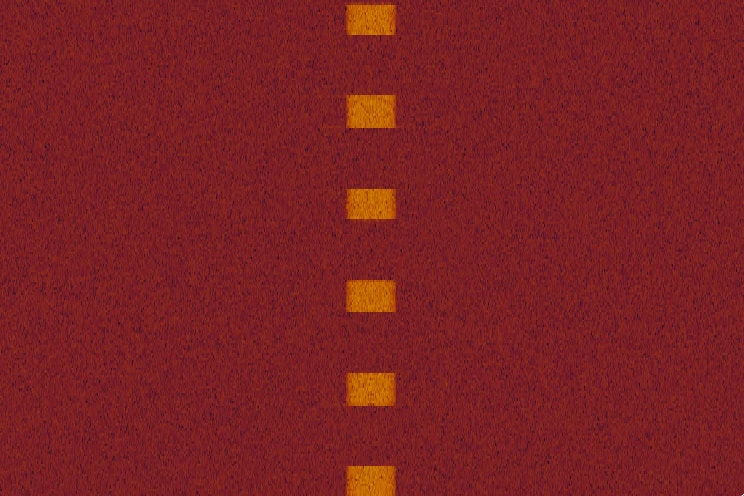}    
       \caption{\small Node failure}
        \end{subfigure}~
        \begin{subfigure}[t]{.2\textwidth}      
        \centering
       \includegraphics[trim= 0.0cm 0.0cm 0.0cm 0.0cm, clip=true, width=\textwidth, height=2.4cm]{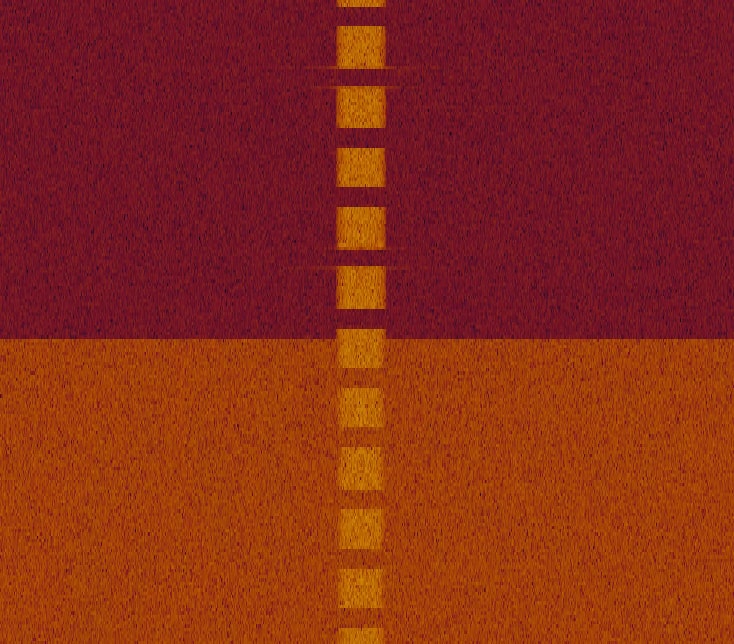}    
       \caption{\small Barrage jamming}
        \end{subfigure}~
        \begin{subfigure}[t]{.2\textwidth}      
        \centering
       \includegraphics[trim= 0.0cm 0.0cm 0.0cm 0.0cm, clip=true, width=\textwidth, height=2.4cm]{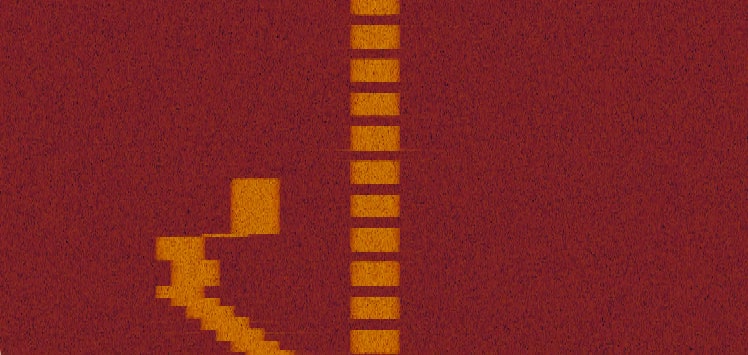}    
       \caption{\small Sweep jamming}
        \end{subfigure}~
        \begin{subfigure}[t]{.2\textwidth}      
        \centering
       \includegraphics[trim= 0.0cm 0.0cm 0.0cm 0.0cm, clip=true, width=\textwidth, height=2.4cm]{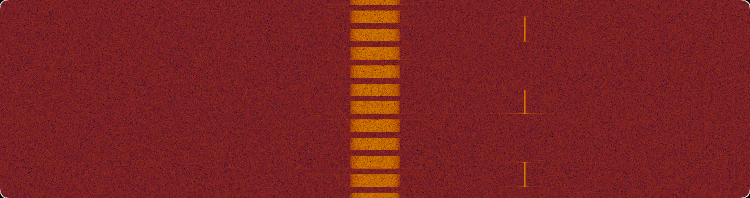}    
       \caption{\small Tone jamming}
        \end{subfigure}
        \caption{\small Example spectrogram image frames from the dataset showing different anomalies used for evaluating our method.}
        \label{fig:anomalies}
\end{figure*}

\begin{figure*}[ht]
\centering
\begin{subfigure}[t]{.5\textwidth}
\begin{subfigure}[t]{.5\textwidth}
\centering
\includegraphics[trim= 0.75cm 0.0cm 1.0cm 0.0cm, clip=true, width=\textwidth]{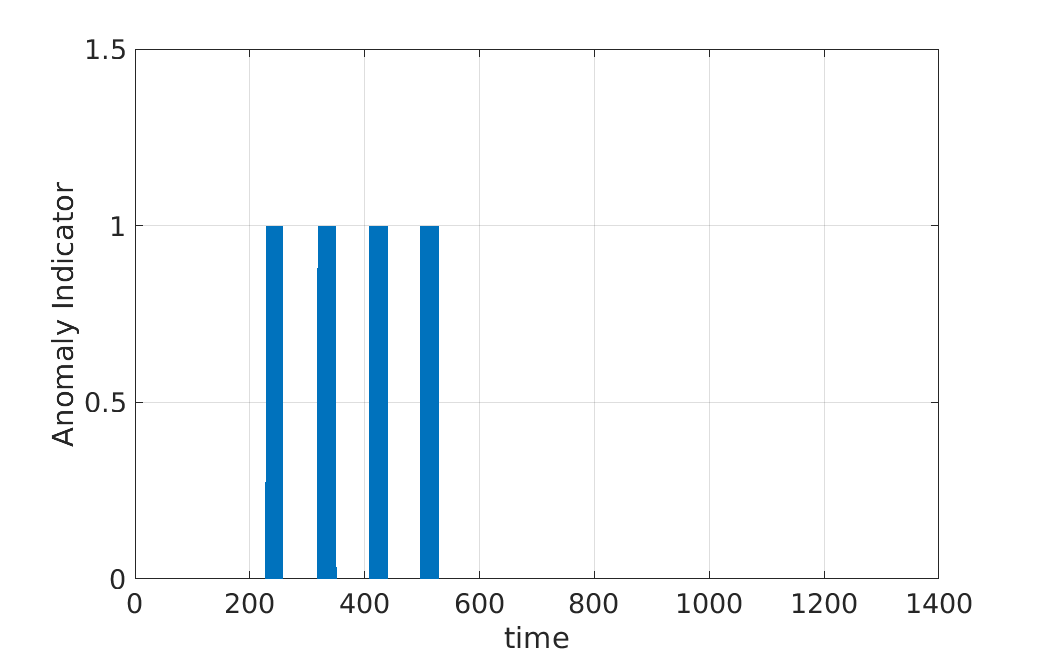}
\end{subfigure}~
\begin{subfigure}[t]{.5\textwidth}
\centering
\includegraphics[trim= 0.75cm 0.0cm 1.0cm 0.0cm, clip=true, width=\textwidth]{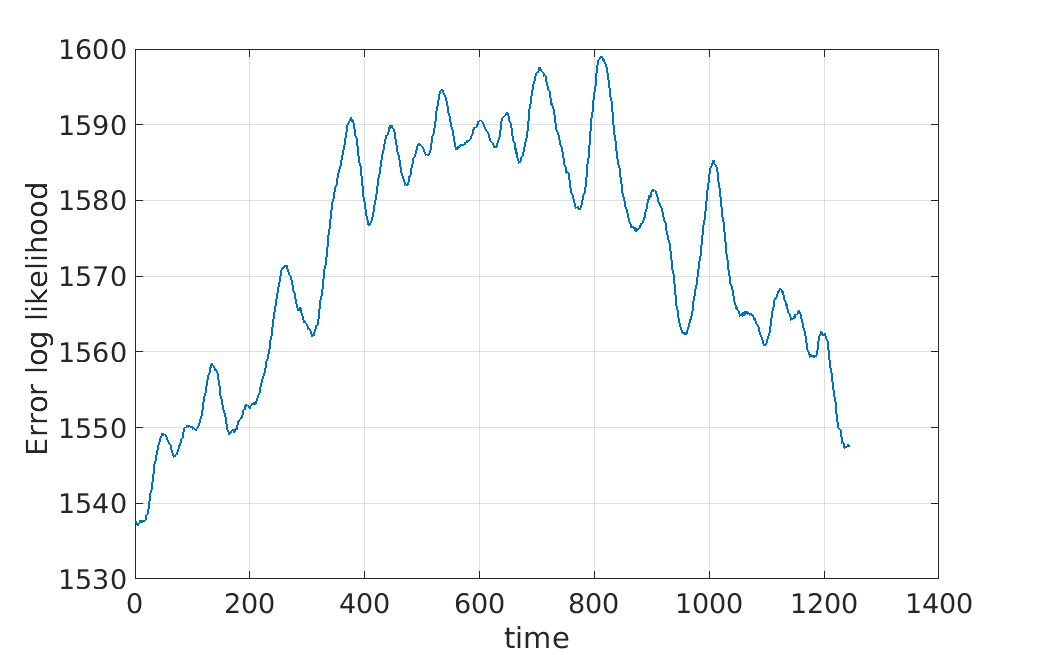}
\end{subfigure}
\caption{\small Single node chirping}
\end{subfigure}~
\begin{subfigure}[t]{.5\textwidth}
\begin{subfigure}[t]{.5\textwidth}
\centering
\includegraphics[trim= 0.75cm 0.0cm 1.0cm 0.0cm, clip=true, width=\textwidth]{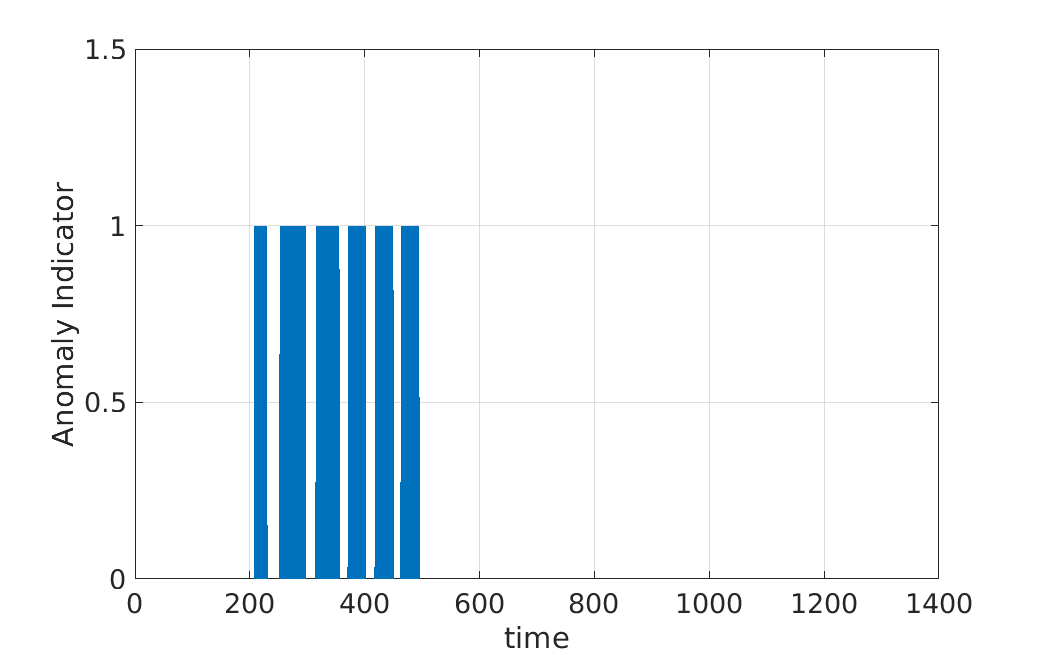}
\end{subfigure}~
\begin{subfigure}[t]{.5\textwidth}
\centering
\includegraphics[trim= 0.75cm 0.0cm 1.0cm 0.0cm, clip=true, width=\textwidth]{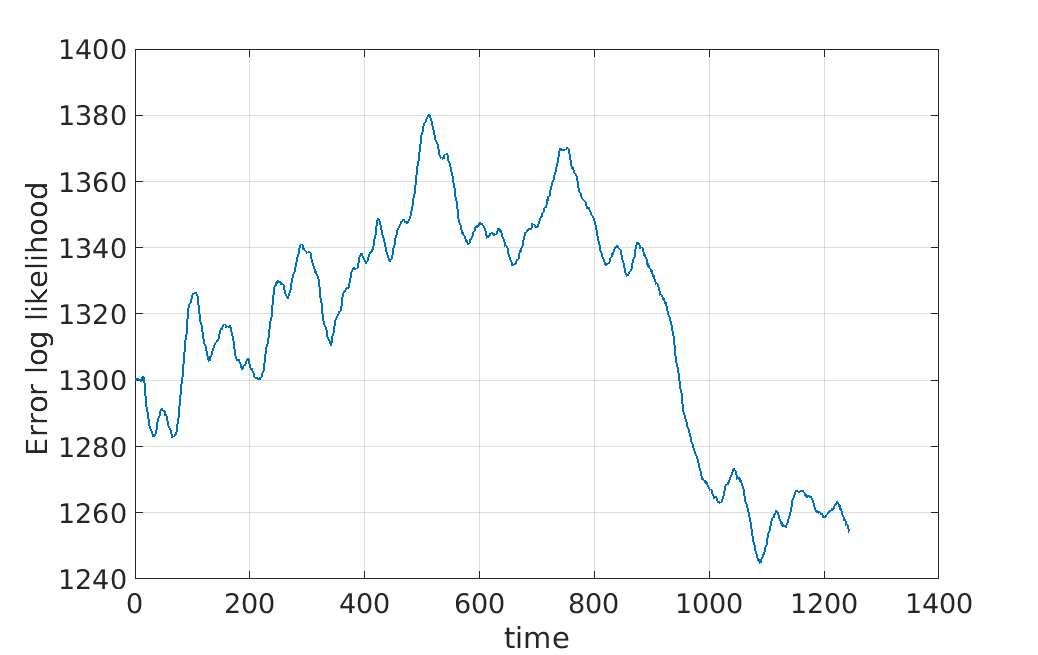}
\end{subfigure}
\caption{\small Two chirping nodes}
\end{subfigure}
\begin{subfigure}[t]{.5\textwidth}
\begin{subfigure}[t]{.5\textwidth}
\centering
\includegraphics[trim= 0.75cm 0.0cm 1.0cm 0.0cm, clip=true, width=\textwidth]{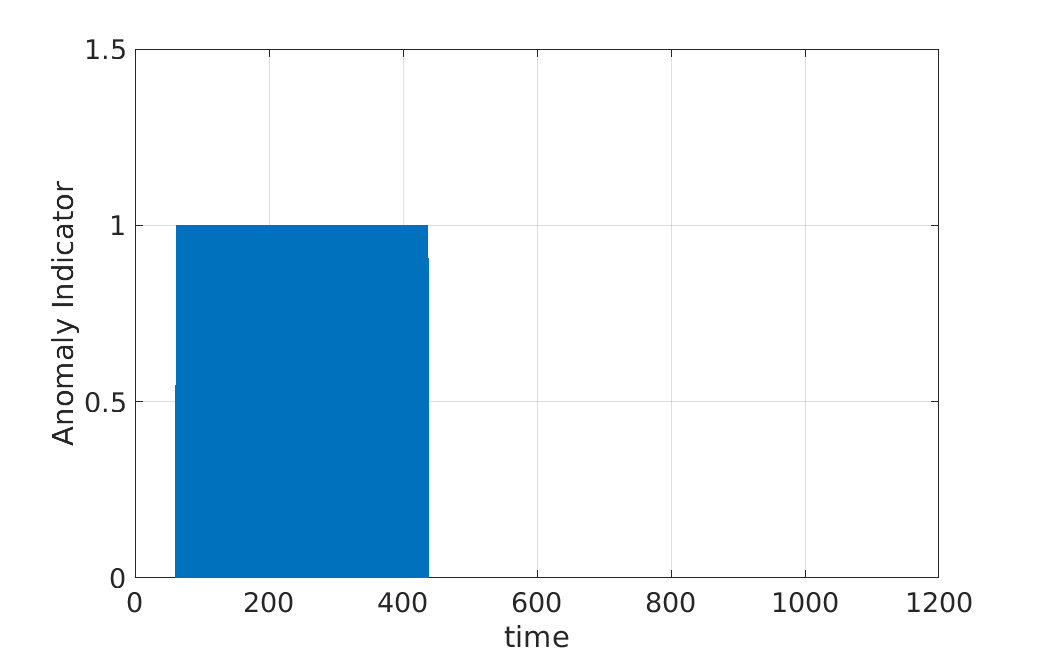}
\end{subfigure}~
\begin{subfigure}[t]{.5\textwidth}
\centering
\includegraphics[trim= 0.75cm 0.0cm 1.0cm 0.0cm, clip=true, width=\textwidth]{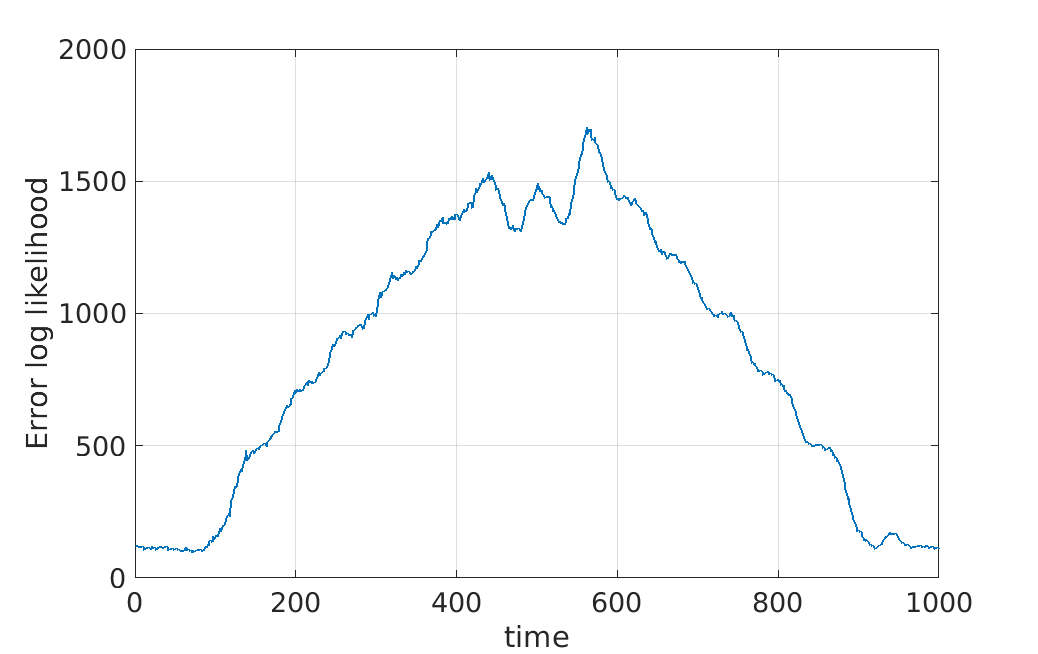}
\end{subfigure}
\caption{\small Barrage Jamming}
\end{subfigure}~
\begin{subfigure}[t]{.5\textwidth}
\begin{subfigure}[t]{.5\textwidth}
\centering
\includegraphics[trim= 0.75cm 0.0cm 1.0cm 0.0cm, clip=true, width=\textwidth]{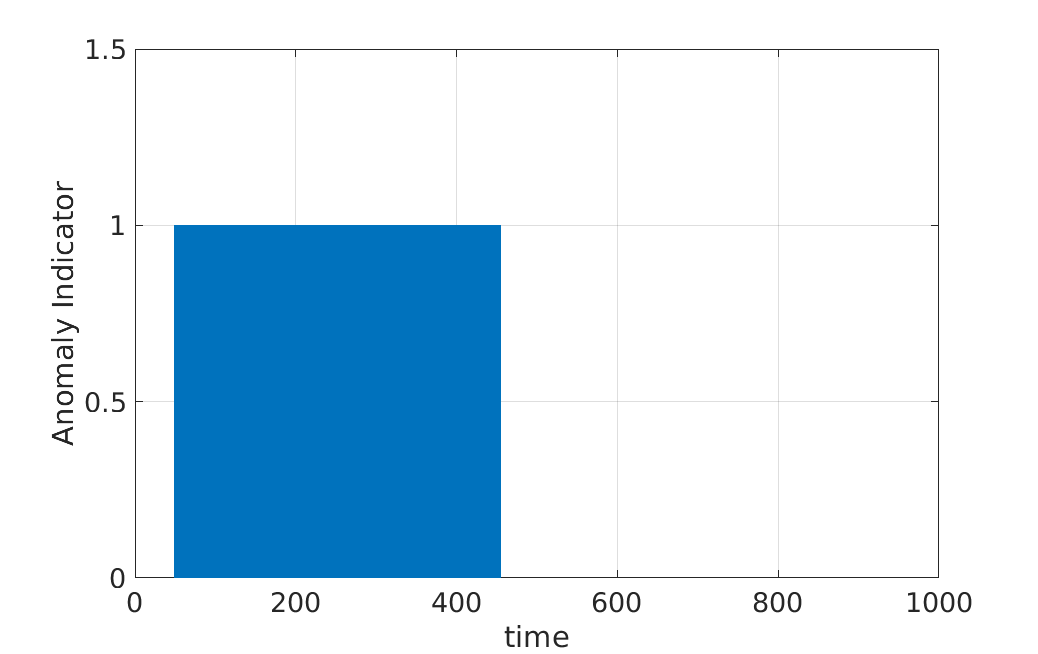}
\end{subfigure}~
\begin{subfigure}[t]{.5\textwidth}
\centering
\includegraphics[trim= 0.75cm 0.0cm 1.0cm 0.0cm, clip=true, width=\textwidth]{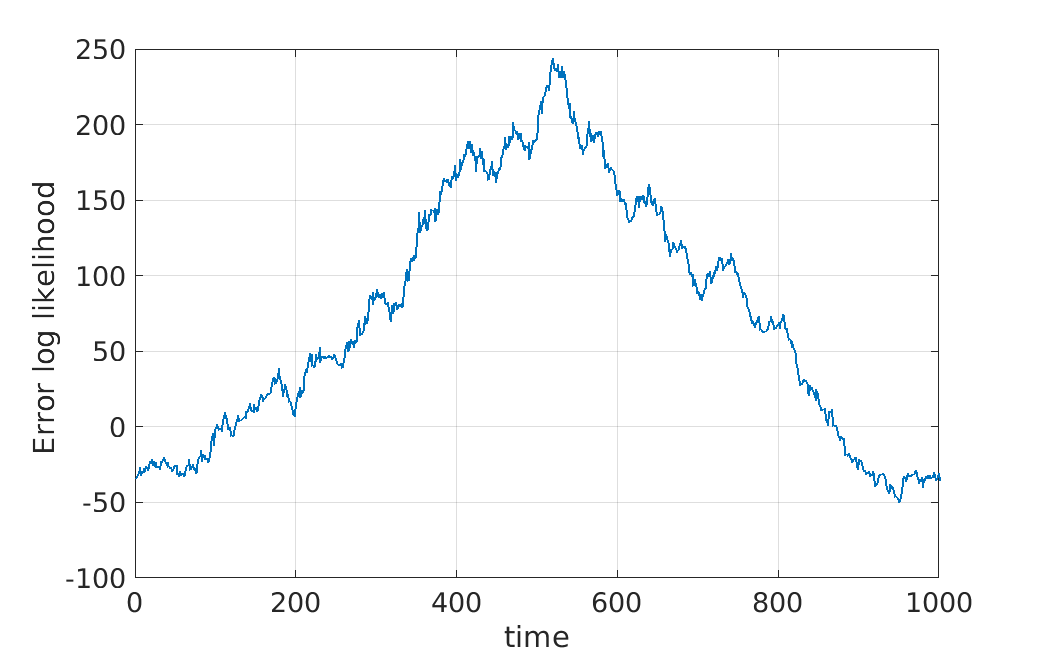}
\end{subfigure}
\caption{\small Sweep Jammer}
\end{subfigure}

\begin{subfigure}[t]{.5\textwidth}
\begin{subfigure}[t]{.5\textwidth}
\centering
\includegraphics[trim= 0.75cm 0.0cm 1.0cm 0.0cm, clip=true, width=\textwidth]{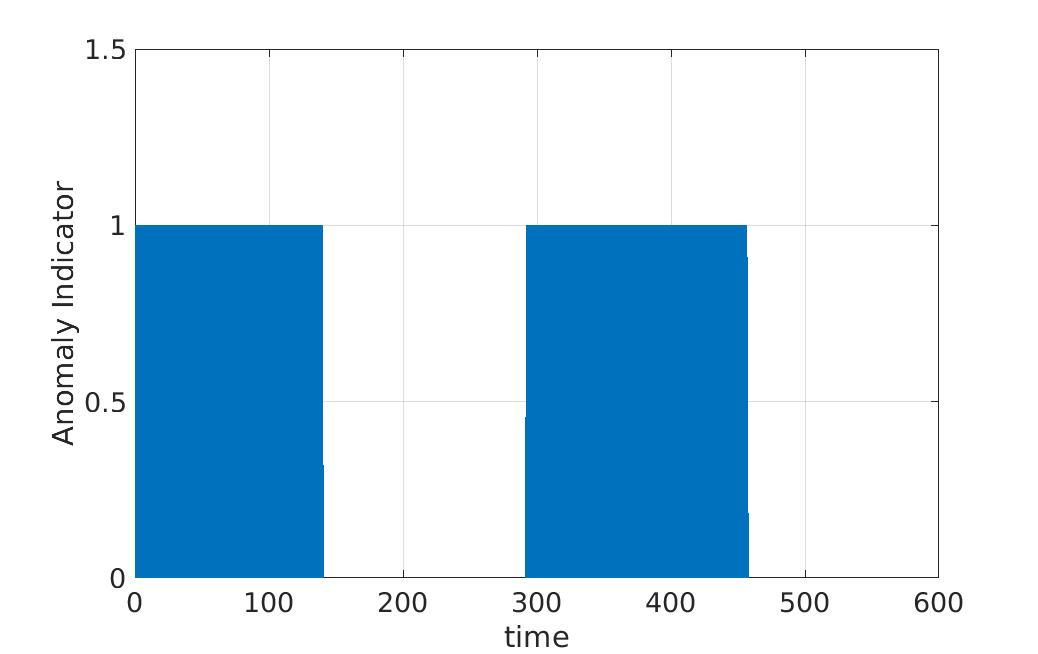}
\end{subfigure}~
\begin{subfigure}[t]{.5\textwidth}
\centering
\includegraphics[trim= 0.75cm 0.0cm 1.0cm 0.0cm, clip=true, width=\textwidth]{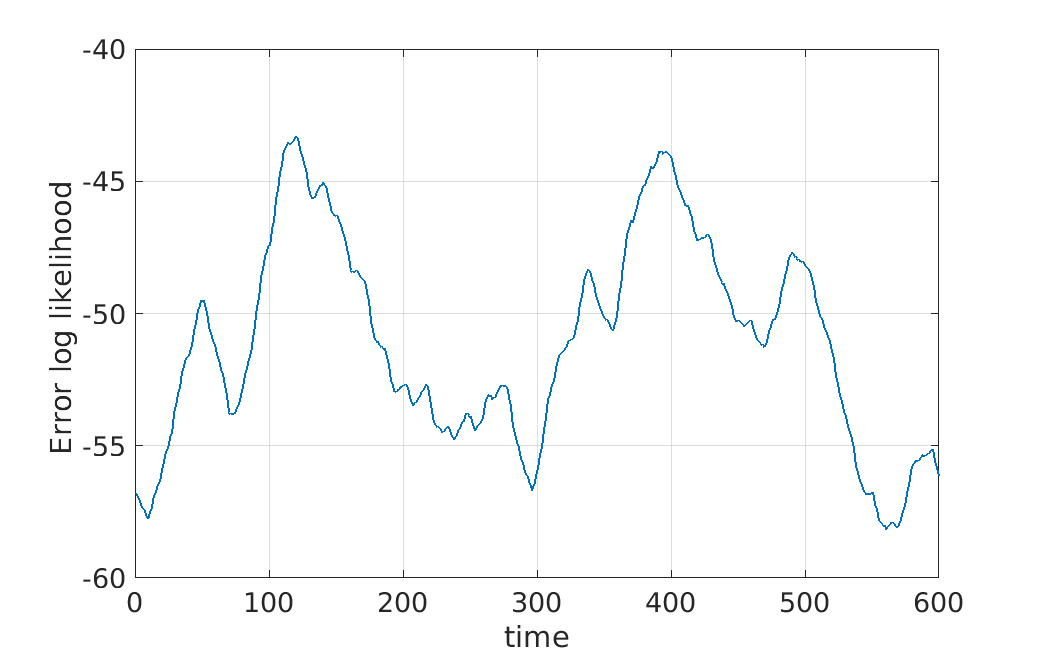}
\end{subfigure}
\caption{\small Spectrum Hijacking}
\end{subfigure}

\caption{\small Performance of our anomaly detector on spectrogram image frames for different anomaly scenarios. For each subfigure, the left side plots the ground truth of anomalous event occurrence over time and the right side plots the detector's error metric.}
\label{fig:respsd}
\end{figure*}
\begin{figure*}
\centering
\begin{subfigure}[t]{.5\textwidth}
\begin{subfigure}[t]{.5\textwidth}
\centering
\includegraphics[trim= 0.75cm 0.0cm 1.0cm 0.0cm, clip=true, width=\textwidth]{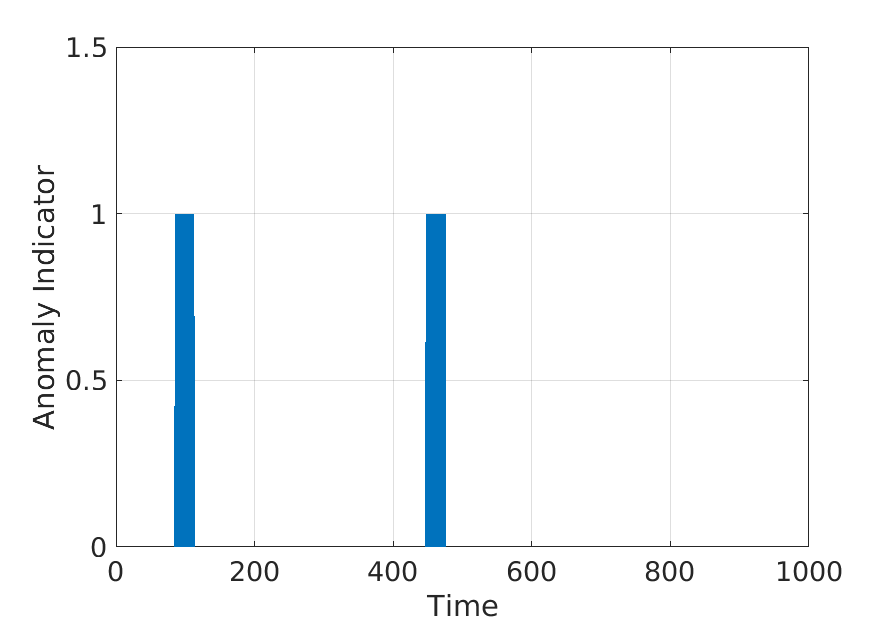}
\end{subfigure}~
\begin{subfigure}[t]{.5\textwidth}
\centering
\includegraphics[trim= 0.75cm 0.0cm 1.0cm 0.0cm, clip=true, width=\textwidth]{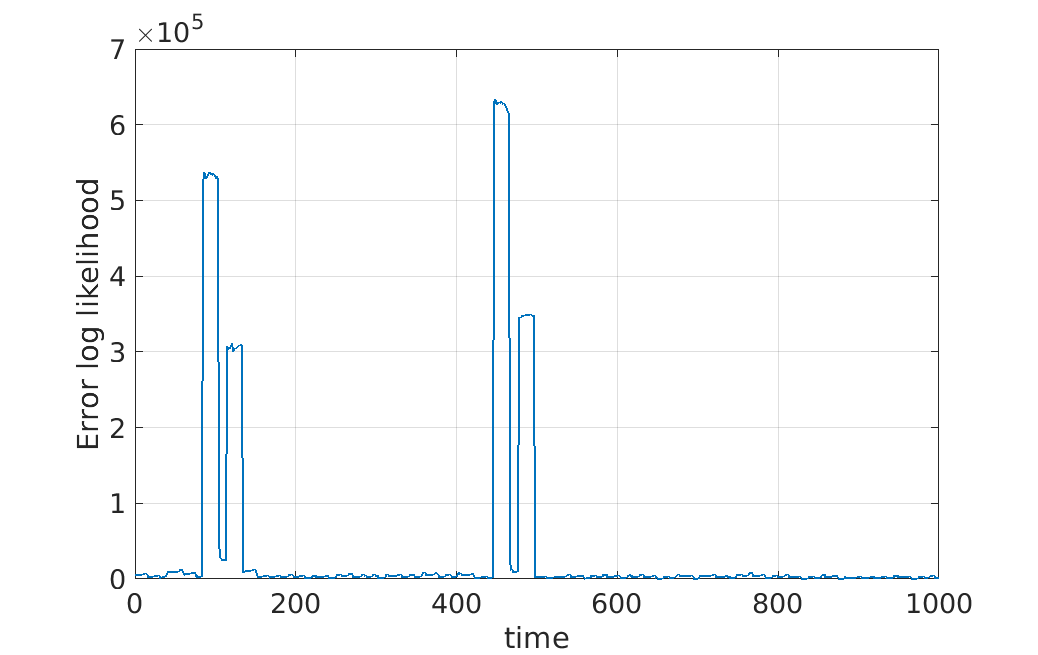}
\end{subfigure}
\caption{Single node chirping}
\end{subfigure}~
\begin{subfigure}[t]{.5\textwidth}
\begin{subfigure}[t]{.5\textwidth}
\centering
\includegraphics[trim= 0.75cm 0.0cm 1.0cm 0.0cm, clip=true, width=\textwidth]{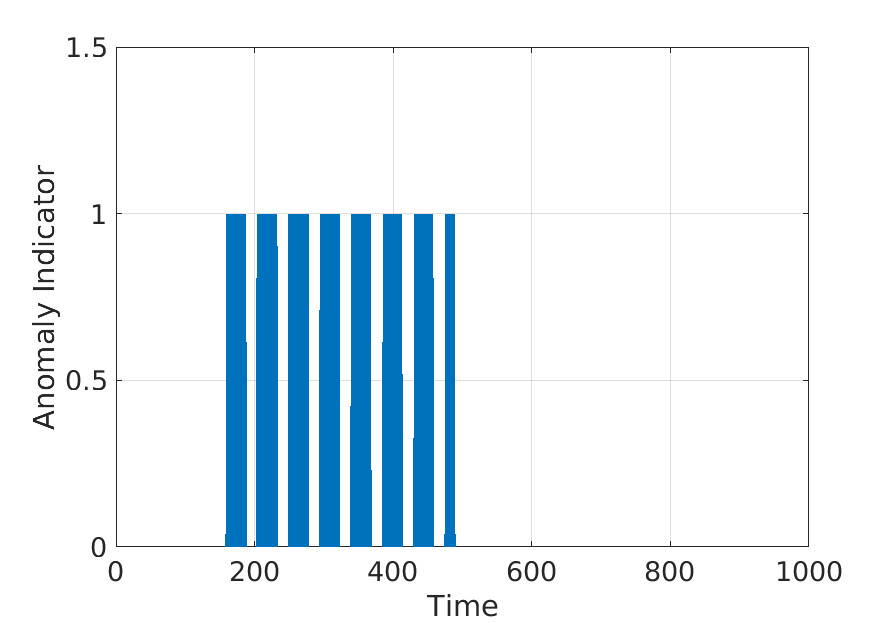}
\end{subfigure}~
\begin{subfigure}[t]{.5\textwidth}
\centering
\includegraphics[trim= 0.75cm 0.0cm 1.0cm 0.0cm, clip=true, width=\textwidth]{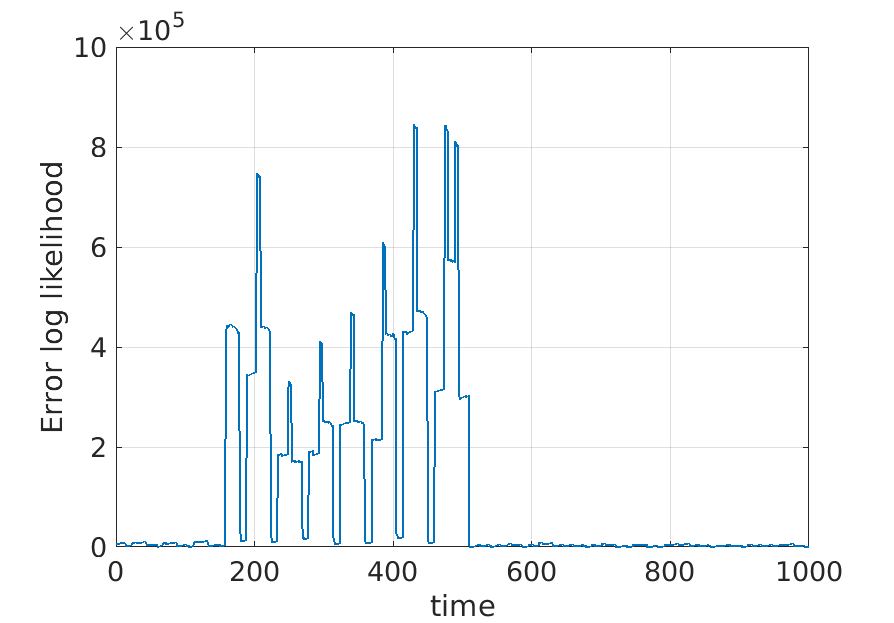}
\end{subfigure}
\caption{Two chirping nodes}
\end{subfigure}
\begin{subfigure}[t]{.5\textwidth}
\begin{subfigure}[t]{.5\textwidth}
\centering
\includegraphics[trim= 0.75cm 0.0cm 1.0cm 0.0cm, clip=true, width=\textwidth]{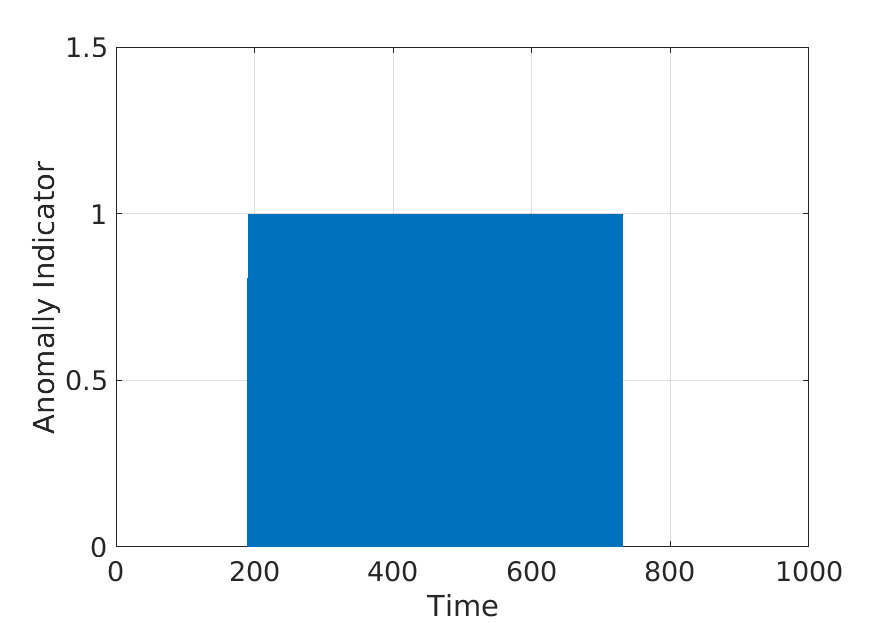}
\end{subfigure}~
\begin{subfigure}[t]{.5\textwidth}
\centering
\includegraphics[trim= 0.75cm 0.0cm 1.0cm 0.0cm, clip=true, width=\textwidth]{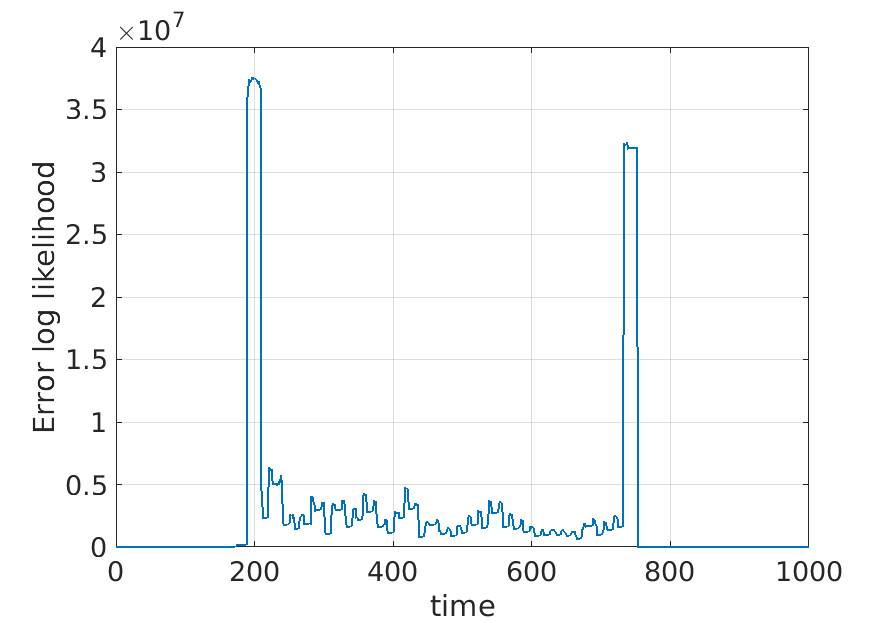}
\end{subfigure}
\caption{Barrage Jamming}
\end{subfigure}~
\begin{subfigure}[t]{.5\textwidth}
\begin{subfigure}[t]{.5\textwidth}
\centering
\includegraphics[trim= 0.75cm 0.0cm 1.0cm 0.0cm, clip=true, width=\textwidth]{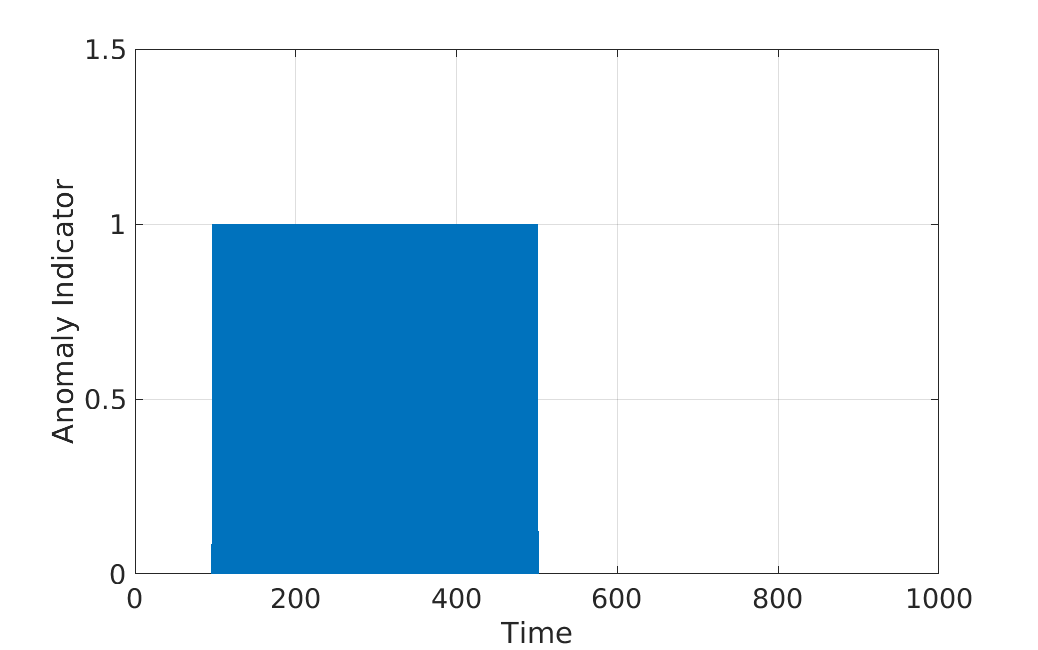}
\end{subfigure}~
\begin{subfigure}[t]{.5\textwidth}
\centering
\includegraphics[trim= 0.75cm 0.0cm 1.0cm 0.0cm, clip=true, width=\textwidth]{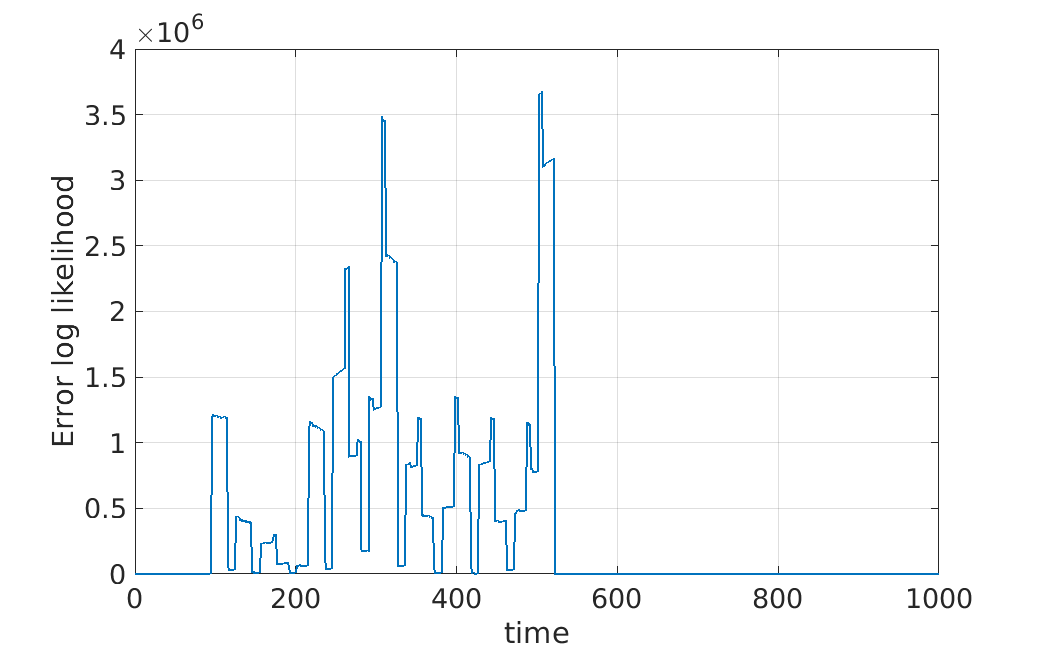}
\end{subfigure}
\caption{Sweep Jammer}
\end{subfigure}
\begin{subfigure}[t]{.5\textwidth}
\begin{subfigure}[t]{.5\textwidth}
\centering
\includegraphics[trim= 0.75cm 0.0cm 1.0cm 0.0cm, clip=true, width=\textwidth]{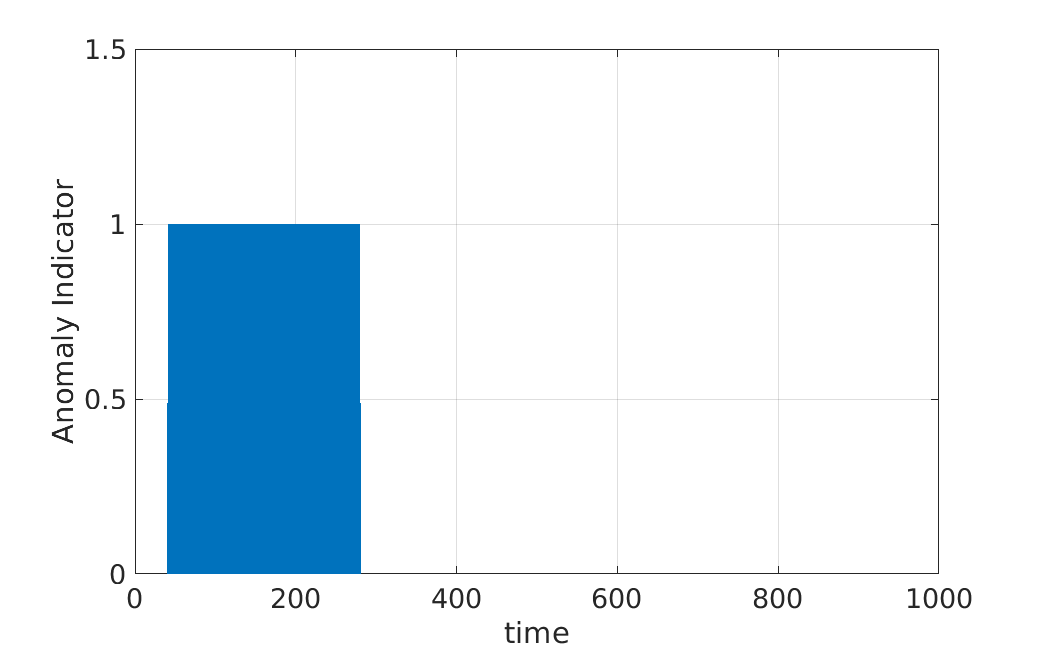}
\end{subfigure}~
\begin{subfigure}[t]{.5\textwidth}
\centering
\includegraphics[trim= 0.75cm 0.0cm 1.0cm 0.0cm, clip=true, width=\textwidth]{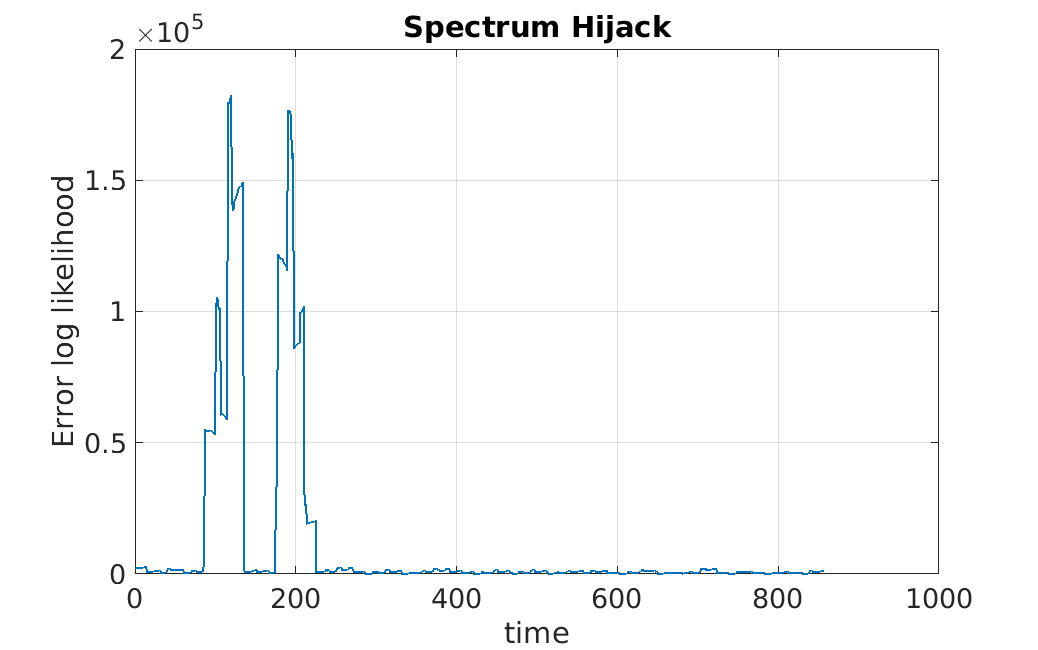}
\end{subfigure}
\caption{Spectrum Hijacking}
\end{subfigure}
\caption{Performance of our anomaly detector on SCF image frames for different anomaly scenarios. For each subfigure, the left side plots the ground truth of anomalous event occurance over time and the right side plots the detector's error metric.}
\label{fig:rescs}
\end{figure*}

\section{Performance Evaluation} \label{sec:results}
\subsection{Implementation Details}
In order to test our method, we implement a three-node network with two source nodes and one sink node. The source nodes share a $500$ kHz channel using time division multiple access (TDMA). The source nodes are transmitting BPSK symbols and the transmitted signal is corrupted by Additive White Gaussian Noise (AWGN). A sample image of spectrum of this system is shown in Fig.~\ref{fig:psd_images}. Our anomaly detector module is located at a sink node which monitors the spectrum and receives the signals transmitted by the network nodes at SNR values $>5$ dB. In order to work at lower SNR, approaches such as \cite{lee1995preprocessing,jafari2012novel} can be used to preprocess the signal. The network implementation and data collection is done using GNU Radio Companion.

In our implementation, we use the two transforms: time-frequency spectrogram and SCF. The spectrogram is computed using the GNU Radio QTGUI waterfall block. The SCF is computed using the FFT accumulation method (FAM) \cite{loomis1991cyclo}, and is implemented in the GNU Radio module \textit{gr-specest} \cite{specest} as
\begin{equation}
S_{xx}^\alpha(n,f) = \frac{1}{N} \sum_{r=1}^N \frac{1}{N'}X_{N'}(n,f+\alpha/2) X_{N'}^*(n,f-\alpha/2). \\
\end{equation}
The FAM algorithm generates a bi-frequency plane, with cycle frequency \(\alpha\) as one axis. The cycle frequency resolution is \(\frac{f_s}{P\times L}\), where \(f_s\) is the sampling rate, $L$ the data chunk size of the first FFT and $P$  the size of the second FFT \cite{loomis1991cyclo}. We use the \textit{gr-inspector} visualization tool to generate SCF images \cite{gnuinspector}.

As mentioned, our anomaly detection module uses the video prediction network \textit{Prednet} \cite{lotter2016deep}, which is a deep predictive coding neural network. It is designed as a multilevel arrangement of local predictors, each of which comprises of an input convolution layer, recurrent representation layer, and an error representation layer. At each level, the recurrent representation layer uses inputs from its own error representation as well as from representation layers from levels above it. Thus the information flow among different levels is top-down, making it equivalent to a generative deconvolutional layer. The network is trained using the back-propagation algorithm. For our simulations, we use a three-level network whose representation layers consist of 48, 96, and 192 channels respectively. The filters of the convolutional layer are of size $(3,3,3).$

Two separate neural networks are tuned and trained for spectrograms and SCF images with dataset which only consist of images corresponding to the normal-behavior of the system. The detectors were trained for 20 and 40 epochs respectively, and the training time for each epoch was 23 s with a IBM Power8 CPU (3.26 GHz, 256 GB memory) and NVIDIA P100 GPU (16 GB memory). The image sequences are evaluated with respect to the trained neural networks. 
\begin{figure*}
\centering
\begin{subfigure}[t]{.5\textwidth}
\centering
\includegraphics[trim= 0.0cm 0.0cm 0.5cm 0.0cm, clip=true, width=\textwidth]{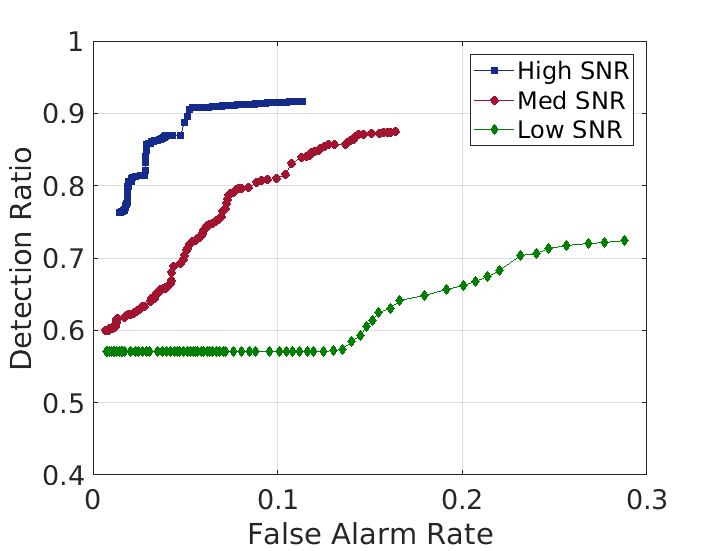}
\caption{\small Detector's performance with spectrogram images.}
\label{fig:overallPSD}
\end{subfigure}~
\begin{subfigure}[t]{.5\textwidth}
\centering
\includegraphics[trim= 0.0cm 0.0cm 0.5cm 0.0cm, clip=true, width=\textwidth]{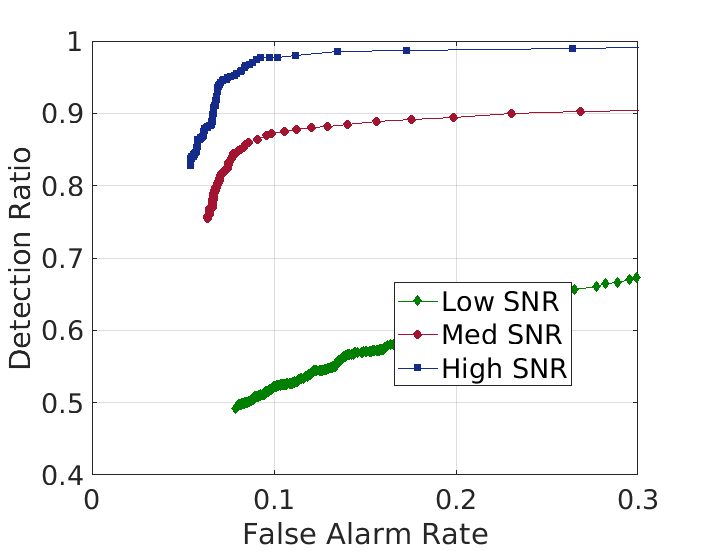}
\caption{\small Detector's performance with SCF images.}
\label{fig:overallCS}
\end{subfigure}
\caption{\small Detection ratio over false alarm rate for the two proposed detectors.}
\end{figure*}

\subsection{Results}
We define several anomalies---chirp, spectrum hijacking, node failure, barrage jamming, sweep jamming, and tone jamming---in our dataset to evaluate the performance of our method. Sample spectrogram images corresponding to these anomalies are shown in Fig.~\ref{fig:anomalies}. We use standard binary classification metrics: detection rate and false alarm rate. Fig.~\ref{fig:respsd} shows the raw output of our anomaly detector when it uses spectrogram data alongside with the anomaly indicator function in the testing dataset. It can be observed that when an anomaly occurs in the network, the average slope of the detector output increases with a high rate. This steady increase in the slope continues while the anomaly stays in the network. After the anomalous event the detector's output gradually goes back to its steady state. Thus, the detector can track the anomaly by observing the slope of its output or error likelihood value, and its detection ratio and false alarm rate is a function of the set threshold for the slope. Fig.~\ref{fig:overallPSD} displays this relationship and was obtained by varying the threshold value for the slope of the error likelihood output of the detector. The figure also shows how the SNR affects the performance of the detector. We can say that accurate detection is possible at high SNR and degrades for lower SNR using the spectrogram-based detector.

Next, we analyze the performance of our detector on the dataset containing SCF images. Fig.~\ref{fig:rescs} shows sample outputs of the SCF image-based detector for five different anomalous RF events. The detector performance can be compared with the ground truth or actual anomaly indicator function plot (shown on the left side in each subfigure). We observe that in most of the cases, the error value obtained by our detector increases almost instantly as the anomaly occurs. This is so because each frame of the input SCF image reflects instantaneous spectral behavior. In order to process this error likelihood curve, we simply need to set a threshold value for the error curve and an anomaly is detected if the error likelihood exceeds this threshold. 
The tradeoff relationship between detection ratio and the false alarm rate is shown in Fig.~\ref{fig:overallCS}. It shows that nearly 100\% detection accuracy can be obtained for high SNR and 90\% for medium SNRs at false alarms rate of 0.15.

If we compare the two detectors, we observe the following:
\begin{enumerate}
\item The spectrogram based detector requires analysis of multiple image frames to decide the presence of an anomaly, whereas the SCF-based detector reacts instantaneously. This difference in behavior of the two detectors can be attributed to the fact that spectrogram image frames contain information for a fixed time window, and the adjacent images in the dataset are highly correlated. Hence, the spectrogram based detector cannot localize anomalies in time domain as accurately as the SCF-based detector.
\item Spectrogram based detectors, on the other hand, can afford to skip processing of adjacent image frames, which makes them useful for monitoring applications when the available computational resources are limited. Example scenarios would be IoT networks or a battery constrained systems, such as WSNs.
\end{enumerate}

\section{Conclusions}\label{sec:conc}
This paper has proposed a ML-based anomaly detection approach for identifying abnormal events in wireless networks. In our method, we monitor real-time wireless signals, process them continuously to obtain power spectral density and SCF video frames. These videos are then passed to the video frame predictor \textit{PredNet}, which iteratively processes each frame to predict the network's behavior in the next frame. Our detector analyzes the deviation of the predicted frame from the actual network behavior to identify the presence of an anomalous event. 

We have trained the video predictor with video frames that correspond to the normal behavior of a network. This way our detection approach can identify unforeseen network anomalies that have different RF signatures. We illustrate this by evaluating our approach against various anomalies, including jamming, spectral hijacking, and chirp signaling. Further, we analyze the detection ratio vs. false alarm rate tradeoff for our detector on two types of spectral datasets: time-frequency plots of power spectral density and cyclostationary profiles of the signal. Because of the high correlation among consecutive spectrogram images, the first detector cannot instantaneously detect an anomaly. With the SCF dataset, on the other hand, the localization of anomalies is more accurate and instantaneous. The performance gain of the SCF-image based detector comes at the cost of higher computational requirement to generate the SCF. 

The proposed anomaly detection methodology can cater demands of networks with variable constraints and requirements. Our approach is scalable to networks with a large number of devices, such as massive machine-type communications in sensor networks, industry automation and the smart grid, to name a few.

We are currently analyzing the implications of different wireless network architectures, consider hyper-parameter searches, evaluate longer runs, larger datasets and additional types of anomalies and combinations of anomalies. More robustness in the detection can be achieved by ML techniques that process multiple transforms of the raw data or preprocessed data.% We will also compare and contrast the performance of our method with existing techniques for one class classification problem.%The other direction for this project could be analysing performance of other prediction networks for this application.   

\section*{ACKNOWLEDGMENT}
This work was partially supported by NSF through grant 1265886 and by the industrial affiliates of the Broadband Wireless Access and Applications Center. The authors would also like to thank Dr. Bert Huang, whose course gave them an opportunity to work on this project.
%\begin{scratch}
%\section{Scratch}
%Note: This section is for putting notes and comments.
% Anomaly detection task of spectrum in wireless communication is quite different from other anomaly detection tasks, mainly reflected in two aspects: 
% (a) the variety of anomaly types makes it impossible to get the label of
% abnormal data. 
% (b) the complexity and the quantity of the electromagnetic environment
% data increase the difficulty of manual feature extraction.
% The failure of current wireless protocols to address these
% vulnerabilities makes intrusion detection for wireless networks extremely important.
% \end{scratch}

\bibliography{bibLoc}
\bibliographystyle{ieeetr}
\end{document}